\documentclass{raa}
\usepackage{natbib}
\usepackage{booktabs}
\usepackage{multirow}
\usepackage{graphicx,times}
\usepackage{amssymb,amsmath}
\usepackage{lineno}
  
\newcommand{\RV}{$\mathrm{RV}$}
\newcommand{\RGC}{$\mathrm{R_{GC}}$}
\newcommand{\LZ}{$\mathrm{L_{Z}}$}
\newcommand{\VP}{$\mathrm{V_{\phi}}$}
\usepackage{cite}

\begin{document}
\title{Search for the metal-weak thick disk from the LAMOST DR5}
\author{T.-S. Yan
  \inst{1,2}
\and J.-R. Shi
  \inst{1,2}
\and H. Tian
  \inst{3}
\and W. Zhang
  \inst{1}
\and B. Zhang
  \inst{4}
}
\institute{Key Laboratory of Optical Astronomy, National Astronomical Observatories, Chinese Academy of Sciences, Beijing 100012, China; 
       {\it sjr@bao.ac.cn}\\
       \and
           School of Astronomy and Space Science, University of Chinese Academy of Sciences, Beijing 100049, China\\
      \and
    Key Laboratory of Space Astronomy and Technology, National Astronomical Observatories, Chinese Academy of Sciences, Beijing 100101, People’s Republic of China;
    {\it tianhao@nao.cas.cn}\\
         \and
           Department of Astronomy, Beijing Normal University, Beijing 100875, China\\
       }
\abstract{ Based on the data release of the Large Sky Area Multi-Object Fiber Spectroscopic Telescope survey (LAMOST DR5) and the \emph{Gaia} Early Data Release 3 (\emph{Gaia} EDR3), 
we construct a sample containing 46,109 giant (log~$\emph{g}$ $\leqslant$ 3.5\,dex) stars with heliocentric distance d $\leqslant$ 4 kpc, and the sample is further divided into two groups of the 
inner ({\RGC} $<$ 8.34 \,kpc) and outer region ({\RGC} $>$ 8.34\,kpc). The {\LZ} distributions of our program stars in the panels with different [Fe/H] and [$\alpha$/Fe] suggest that the thick-disk 
consists of two distinct components with different chemical compositions and kinematic properties. For the inner region, the metal-weak thick disk (MWTD) contributes significantly when 
[$\alpha$/Fe] $>+$0.2\,dex and [Fe/H] $<-$0.8\,dex, while the canonical thick-disk (TD) dominates when [Fe/H]$>-$0.8\,dex. However, MWTD clearly appears only when [$\alpha$/Fe] $>+$0.2\,dex and [Fe/H] $<-$1.2\,dex for the outer region, and its proportion is lower than that of the inner region within the same metallicity. Similar result can be obtained from the {\VP} distribution. The high fraction of MWTD in the inner than that in the outer region  imply that MWTD may form in the inner disk, \textbf{and is an observational evidence about the inside-out disk formation scenario}.
}
   \maketitle
\section{Introduction} \label{sec:introduction}
It is well known that the Milky Way is composed of multiple components, such as the thin-disk, the thick-disk and the halo \citep{An2020,Chiappini1997,vanderKruit2011}, and it provides a unique chance for  studying the formation and evolution of galaxies in detail by analysing the full location, kinematics characteristic and chemical compositions of these stellar components.

As an important component of our Galaxy, the thick-disk has been studied by numerous works  since its discovery 
\citep{BeraldoeSilva2021,DiMatteo2011,DiMatteo2019,Franchini2021,Gilmore1983,Girard2006,Helmi2018,Recio-Blanco2014,Ruchti2011}. The thick-disk is composed primarily of older stars, resulting in 
different chemical composition from the young thin-disk  stars. In the metallicity distribution, the thick-disk exhibits a peak aound [Fe/H]$\sim$ $-$0.6\,dex, and most of the thick-disk stars fall within the 
interval of [Fe/H] from $-$1.0\,dex to $-$0.3\,dex \citep{Reddy2008}. At the mooment, there is no uniform conclusion on the upper and lower bounds of the metallicity distribution for the thick-disk 
stars \citep{Reddy2008}, and some disk-like kinematics stars with very low metallicity (down to [Fe/H] = $-1.7$\,dex, or even lower)  have been found \citep{Carollo2010,Carollo2019,Chiba2000,Norris1985}. 
These metal-poor tail stars are commonly considered belong to the metal-weak thick-disk \citep[MWTD,][]{Beers2014,Carollo2010,Carollo2019,Carollo2021,Ivezic2008,Morrison1990,Reddy2008,Ruchti2011}, 
while the remaining component after removing MWTD from the thick-disk is called as the canonical thick-disk \citep[TD,][]{Carollo2010}.

Due to lack of large-scale sample stars of high-precision parameters, the existence of MWTD remains controversial \citep{Ruchti2010}. \citet{Norris1985} suggested that these stars with [Fe/H]$\leqslant-$1.0\,dex and $e \leqslant$ 0.4 should belong to a population, and \citet{Morrison1990} refered to these  stars of $-$1.6\,dex$<$[Fe/H]$<-$1.0\,dex with kinematical and spatial properties similar to the thick-disk ones as MWTD stars.  It need to be pointed out that the metallicity derived by \citet{Norris1985} and \citet{Morrison1990} is besed on the information from the David Dunlap Observatory (DDO) photometry, and \citet{Twarog1994} found that the photometric metallicity used by \citet{Norris1985} has an $\sim$0.5\,dex offset from the spectral ones for giants near [Fe/H]$\sim-$1.2\,dex. In addition, by comparing the  metallicity obtained from the high-resolution spectra with these from the DDO photometry for giants of  \citet{Norris1985} and \citet{Morrison1990}, \citet{Ryan1995} concluded that most stars with the DDO photometric metallicity of [Fe/H]$<-$1.0\,dex have spectral metallicities of [Fe/H]$>-$1.0\,dex. As a result, these stars identified as belonging to MWTD by \citet{Norris1985} and \citet{Morrison1990}, therefore should belong 
to TD. Although the conclusions from \citet{Twarog1994} and \citet{Ryan1995} cannot rule out the existence of MWTD,  the evidences of MWTD suggested by \citet{Norris1985} and \citet{Morrison1990} become weaker.

Fortunately, the advent of large-scale surveys, such as  the Large Sky Area Multi-Object Fibre Spectroscopic Telescope (LAMOST) survey \citep{Cui2012}, the Sloan Digital Sky Survey \citep[SDSS,][]{York2000}, the Radial Velocity Experiment spectroscopic survey \citep[RAVE,][]{Steinmetz2020a,Steinmetz2020b}, and the \emph{Gaia} mission \citep{GaiaCollaboration2021},  make it possible to perform accurate chemodynamical studies on the stellar populations in our Galaxy \citep{Carollo2019}. The spectra from LAMOST, SDSS and RAVE provide estimation of the stellar atmospheric parameters (effective temperature, surface gravity and metallicity) and radial velocities ({\RV}), and the \emph{Gaia} satellite delivers high-precision astrometric data (position, trigonometric parallax, and proper motions). The combination of parameters obtained from spectrum and astrometry is a powerful tool for studying the properties of stellar component related to the Milky Way by investigating the perspectives of stellar spatial  distribution, chemical composition and kinematics.

In this paper, based on a sample combining spectral information from LAMOST and astrometric data from \emph{Gaia}, we examine MWTD. The paper is organized as follows. In Sect.~\ref{sec:sample}, we describe the selection of our sample  stars.  Sect.~\ref{sec:results} presents the search for MWTD in the kinematic space, and the discussion and conclusions are given in  Sect.~\ref{sect:conclusion}.

\section{Sample}\label{sec:sample}
\subsection{Selection of the Sample Stars}
The sample stars are selected from LAMOST DR5 with a range of  the effective temperature from 4,500\,K to 7,000\,K, and only the giants (log~$\emph{g}$ $\leqslant$ 3.5\,dex) with signal-to-noise (S/N) ratio of their spectra higher than 50 have been considered. This removes all the dwarf stars. The stellar atmosphere parameters and [$\alpha$/Fe] ratios are taken from the recommended values of \citet{Xiang2019} with internal uncertainties  of [Fe/H] and [$\alpha$/Fe] less than 0.07 and 0.05\,dex, respectively, while, the radial velocities ({\RV}s) are adopted from LAMOST DR5 \citep{Luo2015}, and only the objects with a {\RV} uncertainty less than 10\,km\,s$^{-1}$ have been selected. Moreover, to further ensure the accuracy of the metallicity, only the objects with a difference of metallicity between LAMOST DR5 and that of \citet{Xiang2019} less than 0.1\,dex are selected. 

The stellar proper motions (PMs) are obtained from \emph{Gaia} EDR3 \citep{GaiaCollaboration2021} with errors both in right ascension and in declination direction less than 0.2\,mas\,yr$^{-1}$, while the geometric distances  ($\mathrm{\emph{r}_{med}}$) from \citet{Bailer-Jones2021} have been adopted. Our sample stars are confined to d $\leqslant$ 4\,kpc with an uncertainty of ($\mathrm{\emph{r}_{hi}}$$-$$\mathrm{\emph{r}_{lo}}$)/($\mathrm{2\times\emph{r}_{med}}$) smaller than 20\%. Here, $\mathrm{\emph{r}_{lo}}$ and $\mathrm{\emph{r}_{hi}}$ are the 16th and 84th percentiles of the posterior \citep[see][for details]{Bailer-Jones2021}. 

\begin{figure}[ht]
\centering
\includegraphics[scale=0.5]{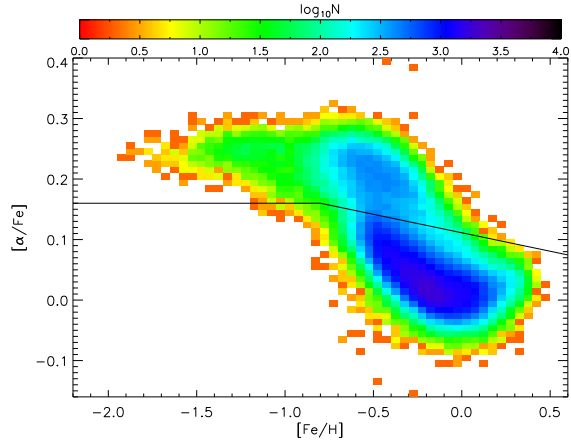}
\caption{Distribution of giant stars in the [$\alpha$/Fe] - [Fe/ H] plane. The black yellow polyline is the criteria for dividing the thin- and thick-disk stars from \citet{Hayden2014}.
\label{fig:feh_afe}}
\end{figure}

After these considerations included, the number of stars is reduced to 211,046, and the distribution of these stars in the [$\alpha$/Fe] - [Fe/H] plane is shown in Fig.~\ref{fig:feh_afe}. Obviously, most of them are belong to the thin-disk, and these thin-disk stars are not important for this study. Therefore, according to the  criteria of \citet{Hayden2014} (see the black yellow polyline in Fig.~\ref{fig:feh_afe}) we remove them from our sample. Finally, there are 46,109 stars in our sample. 

\begin{figure}[ht]
\centering
\includegraphics[scale=0.5]{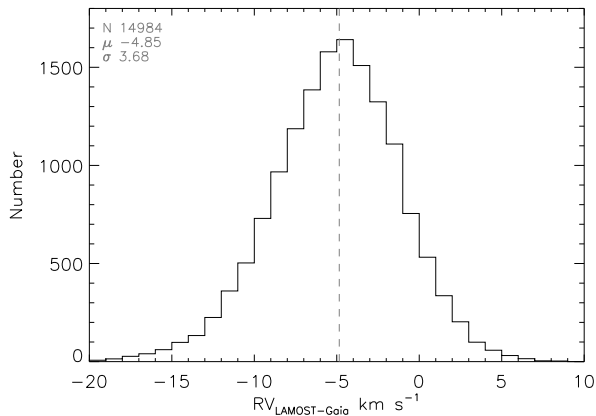}
\caption{Comparison of the LAMOST RVs with those from \emph{Gaia} EDR3, only stars with Gaia RV errors less than 1\,km\,s$^{-1}$ have been used.\label{fig:rv_comp}}
\end{figure}

To avoid the systematic offset of LAMOST RVs \citep{Anguiano2018,Tian2020}, we compare the LAMOST RVs with those  from \emph{Gaia} EDR3 for the common stars, only stars with Gaia RV 
errors less than 1\,km\,s$^{-1}$ have been selected. The offset of LAMOST RV is around $-$4.85\,km\,s$^{-1}$ (Fig.~\ref{fig:rv_comp}), which is very close to those of the previous works 
\citep[$\sim-$5\,km\,s$^{-1}$,][]{Anguiano2018,Tian2020}. Therefore,  a value of 4.85\,km\,s$^{-1}$  is added to the LAMOST RV to compensate for  the offset. 

\subsection{Stellar Kinematic Parameters}\label{sec:parameter}
The full phase-space information, positions (including $\alpha$, $\delta$ and l, b), PMs and distances, combined with {\RV}s,  provide the parameters required to calculate the kinematics. Following \citet{Kordopatis2011},  the Galactic-centerd Cartesian coordinates ($\mathrm{X_{GC}}$, $\mathrm{Y_{GC}}$, Z) of stars are obtained through their distances and Galactic coordinates (l, b), and {\RGC} = $\mathrm{\sqrt{X^2_{GC}+Y^2_{GC}}}$ representes the distance to the Galactic center in planar radial coordinate. In our calculation, the solar position of (X$_{\odot}$, Y$_{\odot}$, Z$_{\odot}=($8.34\,kpc, 0, 0) is adopted \citep{Reid2014}. The galactocentric velocities $\mathrm{V_{R}}$ and {\VP} are also calculated, and $\mathrm{V_{R}}$ and {\VP} are defined as positive with increasing {\RGC} and $\phi$, respectively.  The Galactic rotation velocity of the Local Standard of Rest (LSR) is adopted as V$_{c}$ = 240\,km\,s$^{-1}$ \citep{Reid2014},  and the solar motion with respected to LSR  of ($\mathrm{U_{\odot}, V_{\odot}, W_{\odot}})=$ (11.1, 12.24, 7.25)\,km\,s$^{-1}$, is taken from \citet{Schonrich2010}. The Z-axis angular momentum is derived using the relationship of {\LZ} = {\RGC}$\times${\VP}.

\section{Results}\label{sec:results}
Although MWTD overlaps TD and the halo in parameter space, it is still possible to trace the footprint of MWTD with the help of their distribution in the metallicities, kinematics and dynamics 
\citep{An2020,An2021,Carollo2010,Carollo2019,Carollo2021,Cordoni2021,Kordopatis2013}.

\begin{figure}[ht]
\centering
\includegraphics[scale=0.5]{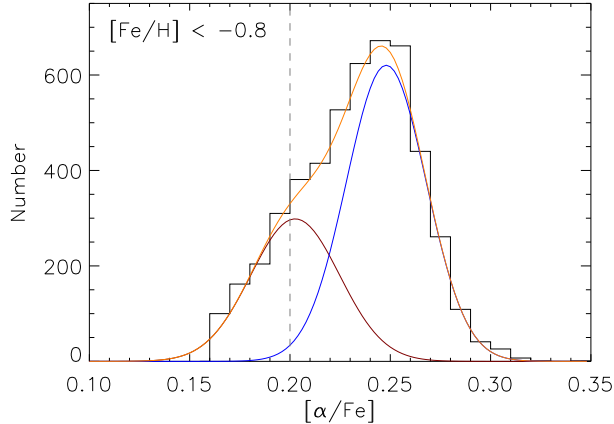}
\caption{[$\alpha$/Fe] distribution of the sample stars with [Fe/H] $<-$0.8\,dex. The distribution can be well fitted with two Gaussian components. The vertical dotted line is labeled as [$\alpha$/Fe] $=+$0.2\,dex.\label{fig:afe}}
\end{figure}

It is noted that the fraction of metal-poor stars is lower \citep[see Fig. 8 of][]{Miranda2016}, and \citet{Carollo2010} pointed out that the significant contribution of MWTD is  within $-$1.8\,dex $<$ [Fe/H] $<-$0.8\,dex, and possibly up to $\sim-$0.7\,dex. Therefore, we divide our sample stars into four metallicity intervals of [Fe/H] $<-$1.2\,dex, $-$1.2\,dex $<$ [Fe/H] $<-$0.8\,dex, $-$0.8\,dex $<$ [Fe/H] $<-$0.4\,dex and [Fe/H] $>-$0.4\,dex. Moreover, \citet{Carollo2019} suggested that the MWTD stars tend to have higher [$\alpha$/Fe] ratios than those of TD,  which means it will improve the search efficiency of MWTD when the sample is decomposed in [$\alpha$/Fe] space. In Fig.~\ref{fig:afe} we plot the [$\alpha$/Fe] distribution of our sample stars of [Fe/H] $<-$0.8\,dex, and the distribution can be well fitted with two Gaussian functions. According to their [$\alpha$/Fe] distribution, the sample is further divided into high ([$\alpha$/Fe] $>+$0.2\,dex)  and low $\alpha$ ([$\alpha$/Fe] $<+$0.2\,dex) intervals.

In order to compare the differences due to different {\RGC} \citep[e.g.,][]{An2021}, we present our discussions on our program stars with the inner ({\RGC} $<$ 8.34, 23,664 stars) and outer ({\RGC} $>$ 8.34\,kpc, 22,445 stars) regions. 

\subsection{The {\LZ} distributions}

\begin{figure*}[ht]
\centering
\includegraphics[scale=1.6]{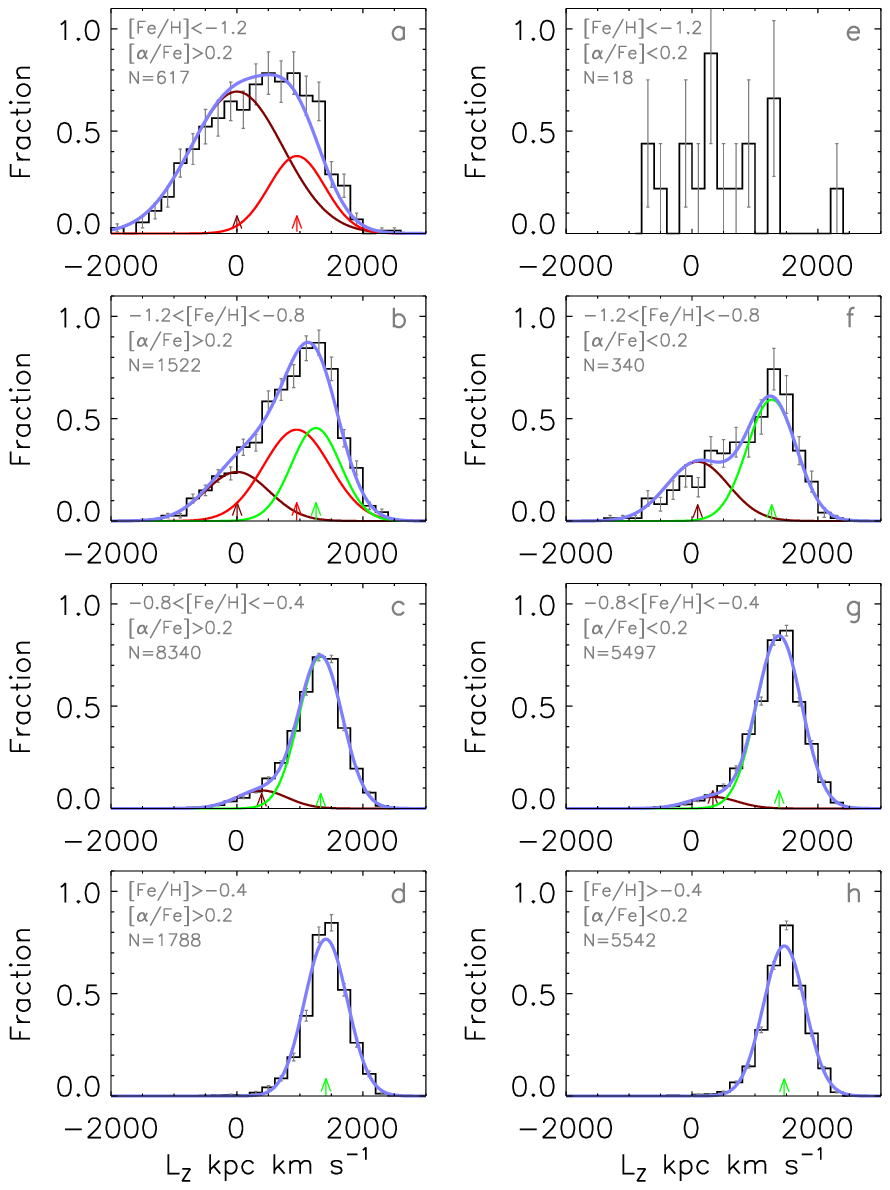}
\caption{The {\LZ} distributions of the inner region stars in slices of [Fe/H] and [$\alpha$/Fe]. The distribution can be  best-fitted by one to three Gaussian(s), and each represents one component:  the halo (brown), MWTD (red) and TD (green). The black histogram is the sum of all stars, while the light blue solid line represents the sum of the fit components. \label{fig:lz_inner}}
\end{figure*}

Following \citet{Carollo2019}, we focus on the {\LZ} distribution. In Fig.~\ref{fig:lz_inner}, we plot the {\LZ} distribution for the inner region stars ({\RGC} $<$ 8.34) in slices of [Fe/H] and [$\alpha$/Fe]. We apply multi-Gaussians to fit the {\LZ} distribution in each panel, and the fitting parameters of each component including the fraction, {\LZ} and scatter ($\mathrm{\sigma_{LZ}}$) are derived according to the Bayesian method, which can effectively avoid the contamination between different components \citep{Tian2019}. To perform this fitting, we apply \emph{emcee} \citep{ForemanMackey2013} to run a Markov Chain Monte Carlo (MCMC) simulation. We use 100 walkers for 2000 iterations with 1000 burn-in have been used. The median value for each parameter is adopted, and the difference between the median value and the 16\% and 84\% values are used as the upper and lower uncertainties, respectively. Fig 5 shows the MCMC result of panel b in Fig.~\ref{fig:lz_inner}. 

The best-fit parameters and uncertainties of the MCMC simulations for all the subsamples are listed in Table~\ref{tab:lz_inner}. When the fraction is not a free parameter during the MCMC simulation (one in each panel),  its uncertainty is marked as 0 (Table~\ref{tab:lz_inner}). The fitted distributions are represented with  brown, red, green and blue lines for the halo, MWTD, TD and total, respectively, in Fig.~\ref{fig:lz_inner}.

\begin{figure*}[ht]
\centering
\includegraphics[scale=0.35]{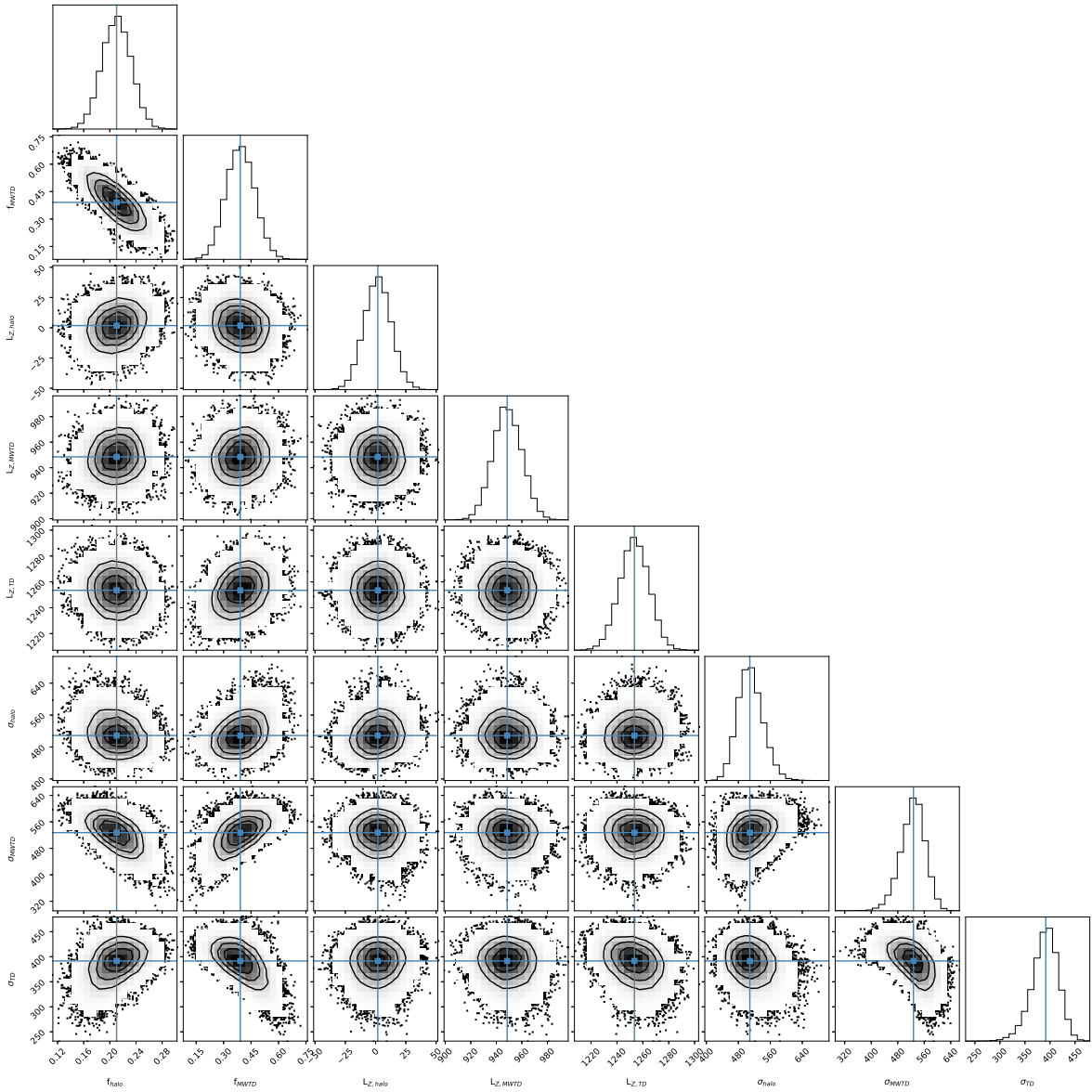}
\caption{The MCMC simulation results of the panel b of Fig. \ref{fig:lz_inner}. The adopted median values of the parameters are represented by the solid blue lines.\label{fig:mcmc}}
\end{figure*}
\begin{table*}[]
\tiny
\centering
\setlength{\tabcolsep}{1.6pt}
\caption{The parameters of the best-fit components in the {\LZ} distributions for the inner region stars in slices of [Fe/H] and [$\alpha$/Fe]. The units of the {\LZ} peak and scatter ($\sigma_{\mathrm{LZ}}$) are in kpc\,km s$^{-1}$.\label{tab:lz_inner}}
\begin{tabular}{ccccccccccccccc}
\toprule
\multirow{3}{*}{[$\alpha$/Fe]} & \multirow{3}{*}{Component} & \multicolumn{3}{c}{[Fe/H]$<$$-$1.2\,dex} & 
\multicolumn{3}{c}{$-$1.2\,dex$<$[Fe/H]$<$$-$0.8\,dex} & \multicolumn{3}{c}{$-$0.8\,dex$<$[Fe/H]$<$$-$0.4\,dex} &
\multicolumn{3}{c}{[Fe/H]$>$$-$0.4\,dex} \\
\cmidrule(r){3-5} \cmidrule(r){6-8} \cmidrule(r){9-11} \cmidrule(r){12-14}
 & &  fraction &  $\mathrm{L_{Z}}$ peak & $\sigma_{\mathrm{LZ}}$ 
&  fraction & $\mathrm{L_{Z}}$ peak & $\sigma_{\mathrm{LZ}}$
&  fraction &  $\mathrm{L_{Z}}$ peak & $\sigma_{\mathrm{LZ}}$ 
&  fraction &  $\mathrm{L_{Z}}$ peak & $\sigma_{\mathrm{LZ}}$    \\
\midrule
\multirow{3}{*}{$>0.2$} & TD & & &  & 0.40$\pm$0 & 1253.5$\pm$11.3 & 391.6$\pm$26.4 & 0.89$\pm$0 & 1328.2$\pm$7.1 & 346.1$\pm$5.0 & 1.00$\pm$0 & 1412.0$\pm$8.0 & 341.3$\pm$5.9 \\
              & MWTD  & 0.35$\pm$0 & 950.7$\pm$7.0 & 444.9$\pm$35.4 & 0.39$\pm$0.08 & 948.5$\pm$11.2 & 527.5$\pm$37.0 \\
              & the halo  & 0.65$\pm$0.03 & -0.2$\pm$7.1 & 733.6$\pm$33.6 & 0.21$\pm$0.02 & 1.8$\pm$11.2 & 508.8$\pm$31.5 & 0.11$\pm$0.01 & 391.6$\pm$41.5 & 405.2$\pm$18.8 \\
\midrule
\multirow{2}{*}{$<0.2$} & TD & & &  & 0.67$\pm$0 & 1265.3$\pm$40.8 & 390.0$\pm$33.7 & 0.94$\pm$0 & 1383.1$\pm$7.2 & 356.2$\pm$5.5 & 1.00$\pm$0 & 1464.0$\pm$4.5 & 331.6$\pm$3.3 \\
              & the halo & & &  & 0.33$\pm$0.05 & 91.2$\pm$61.4 & 490.6$\pm$50.1 & 0.06$\pm$0.01 & 329.6$\pm$48.1 & 384.9$\pm$23.8 \\
\midrule
\end{tabular}
\end{table*}

It can be seen from  Fig.~\ref{fig:lz_inner} that the three  low $\alpha$ ([$\alpha$/Fe] $<+$0.20\,dex) panels (f, g and h) are dominated by the TD stars, and their {\LZ} peaks gradually decrease with decreasing metallicity. 
When [Fe/H] $<-$0.4\,dex, the halo exists (panels f and g), and it accounts for about one third when $-$1.2\,dex $<$ [Fe/H] $<-$0.8\,dex (panel f).

For the four high $\alpha$ panels, the {\LZ} distributions vary regularly with the metallicity. The TD stars dominate both in panels c and d, and the {\LZ} distribution of these two panels are very similar to those of panels g 
and h. The typical {\LZ} peak value is around 1350\,kpc\,km\,s$^{-1}$.

We note that there are three components in panel b, and one of them has a {\LZ} peak of 948.5$\pm$11.2\,kpc\,km\,s$^{-1}$ with a scatter of 527.5$\pm$37.0\,kpc\,km\,s$^{-1}$, this value is lower than that of TD (1253.5$\pm$11.3\,kpc\,km\,s$^{-1}$), while, it is higher than that of the halo (1.8$\pm$11.2\,kpc\,km s$^{-1}$). Which indicates that this component is dynamically different from those of the halo and TD, and it is most likely to be MWTD \citep{An2020,Carollo2010,Naidu2020}. The difference of the {\LZ} peak between MWTD and TD is around 300\,kpc\,km\,s$^{-1}$, which is larger than that of the fitting error. Our results present that the fraction of MWTD in this panel is around 0.39, while the fractions of the halo and TD are $\sim$0.21 and $\sim$0.40, respectively.

For the more metal-poor panel a, although there are only two components, we notice that there is a component with a {\LZ} peak of 950.7$\pm$7.0\,kpc\,km\,s$^{-1}$ with a scatter of 444.9$\pm$35.4\,kpc\,km\,s$^{-1}$.  This {\LZ} peak value is very close to that of MWTD in panel b, which indicates this component is also MWTD. The contribution of MWTD in this panel is about one third (0.35). The remaining contribution in the panel accounts 
for about two-thirds, and they comes from the halo of a {\LZ} peak of $-$0.2$\pm$7.1\,kpc\,km\,s$^{-1}$ with a sctter of 733.6$\pm$33.6\,kpc\,km\,s$^{-1}$. 

\begin{figure*}[ht]
\centering
\includegraphics[scale=1.6]{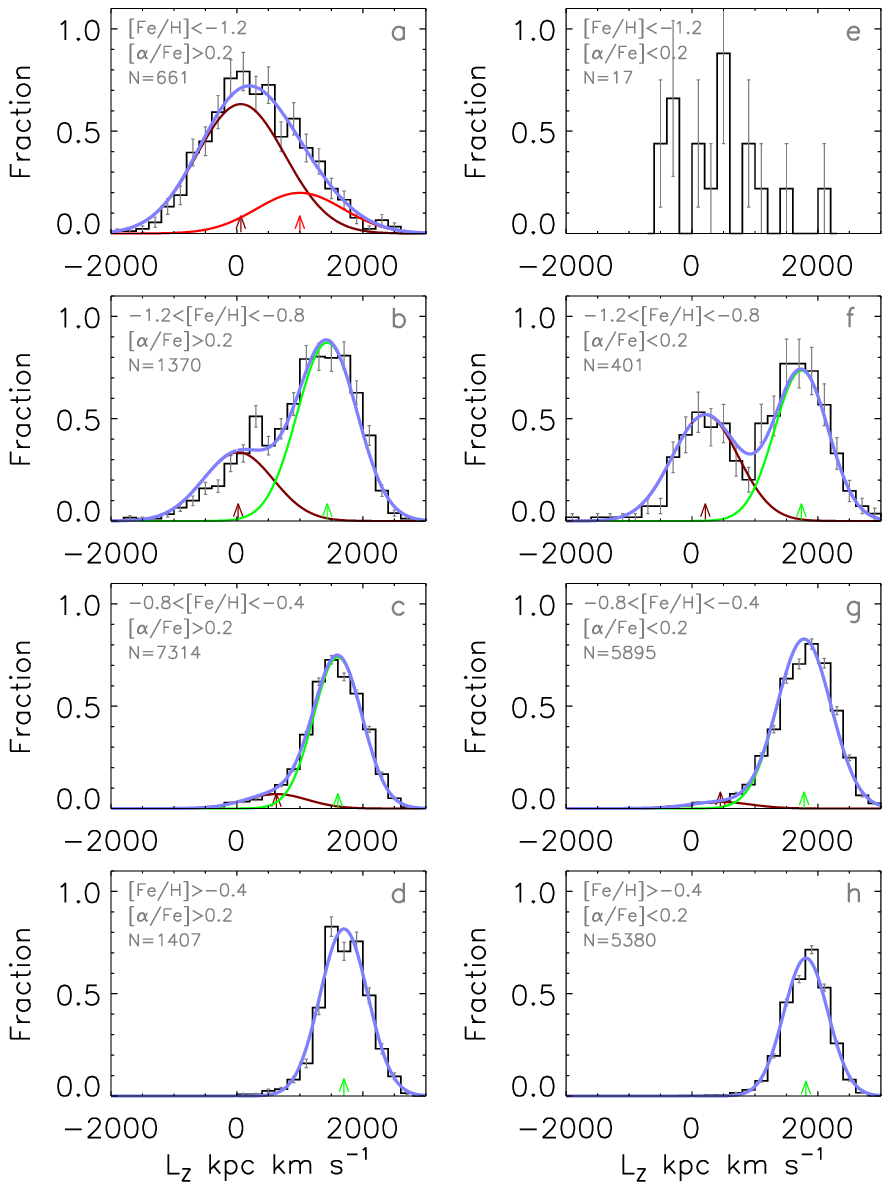}
\caption{Similar to Fig.~\ref{fig:lz_inner} but for the {\LZ} Distribution of outer region stars.\label{fig:lz_outer}}
\end{figure*}

Fig.~\ref{fig:lz_outer}  presents the {\LZ} distributions of the outer region stars ({\RGC} $>$ 8.34\,kpc, black histogram) in slices of [Fe/H] and [$\alpha$/Fe]. Similar to the inner region, we apply \emph{emcee} to perform a MCMC simulation to determine the components contained of each panel for the outer region stars. The best-fit components obtained through the simulation are superimposed on the histogram of Fig. \ref{fig:lz_outer} with color curves, and the best-fit parameters are listed in Table \ref{tab:lz_outer}. 

\begin{table*}[]
\tiny
\centering
\setlength{\tabcolsep}{1.6pt}
\caption{Similar to Table~\ref{tab:lz_inner}, but for the outer region stars.\label{tab:lz_outer}}
\begin{tabular}{ccccccccccccccc}
\toprule
\multirow{3}{*}{[$\alpha$/Fe]} & \multirow{3}{*}{Component} & \multicolumn{3}{c}{[Fe/H]$<$$-$1.2} & 
\multicolumn{3}{c}{$-$1.2$<$[Fe/H]$<$$-$0.8} & \multicolumn{3}{c}{$-$0.8$<$[Fe/H]$<$$-$0.4} &
\multicolumn{3}{c}{[Fe/H]$>$$-$0.4} \\
\cmidrule(r){3-5} \cmidrule(r){6-8} \cmidrule(r){9-11} \cmidrule(r){12-14}
 & &  fraction &  $\mathrm{L_{Z}}$ peak & $\sigma_{\mathrm{LZ}}$ 
&  fraction & $\mathrm{L_{Z}}$ peak & $\sigma_{\mathrm{LZ}}$
&  fraction &  $\mathrm{L_{Z}}$ peak & $\sigma_{\mathrm{LZ}}$ 
&  fraction &  $\mathrm{L_{Z}}$ peak & $\sigma_{\mathrm{LZ}}$    \\
\midrule
\multirow{3}{*}{$>0.2$} & TD & & &  & 0.72$\pm$0 & 1432.5$\pm$12.8 & 475.6$\pm$15.0 & 0.91$\pm$0 & 1600.4$\pm$10.5 & 386.8$\pm$6.4 & 1.00$\pm$0 & 1697.4$\pm$10.2 & 378.7$\pm$7.3 \\
              & MWTD  & 0.24$\pm$0 & 996.9$\pm$22.0 & 687.8$\pm$62.1 \\
              & the halo  & 0.76$\pm$0.04 & 60.8$\pm$21.9 & 701.4$\pm$30.2 & 0.28$\pm$0.02 & 19.1$\pm$15.2 & 555.4$\pm$28.1 & 0.09$\pm$0.02 & 625.3$\pm$136.6 & 494.0$\pm$55.2 \\
\midrule
\multirow{2}{*}{$<0.2$} & TD & & &  & 0.59$\pm$0 & 1736.4$\pm$44.1 & 427.2$\pm$30.1 & 0.96$\pm$0 & 1779.6$\pm$49.8 & 427.9$\pm$14.2 & 1.00$\pm$0 & 1808.0$\pm$4.8 & 350.2$\pm$3.4 \\
              & the halo & & &  & 0.41$\pm$0.04 & 211.3$\pm$86.7 & 525.5$\pm$62.6 & 0.04$\pm$0.01 & 450.4$\pm$30.1 & 485.2$\pm$35.8 \\
\midrule
\end{tabular}
\end{table*}

MWTD has obvious contribution only in panel a of Fig. \ref{fig:lz_outer}. The fraction of MWTD in this panel is 0.24, which is lower than that in the same panel of the inner region. The {\LZ} peak of MWTD in this panel is 996.9$\pm$22.0\,kpc\,km\,s$^{-1}$ with a scatter of 687.8$\pm$62.1\,kpc\,km\,s$^{-1}$, and we note that this scatter is larger than that of the inner ones ($\sim$500\,kpc\,km\,s$^{-1}$). The reason is most likely due to the fraction of MWTD in the outer region is lower than that in the inner one.

\subsection{The {\VP} distributions}

\begin{figure*}[ht]
\centering
\includegraphics[scale=1.6]{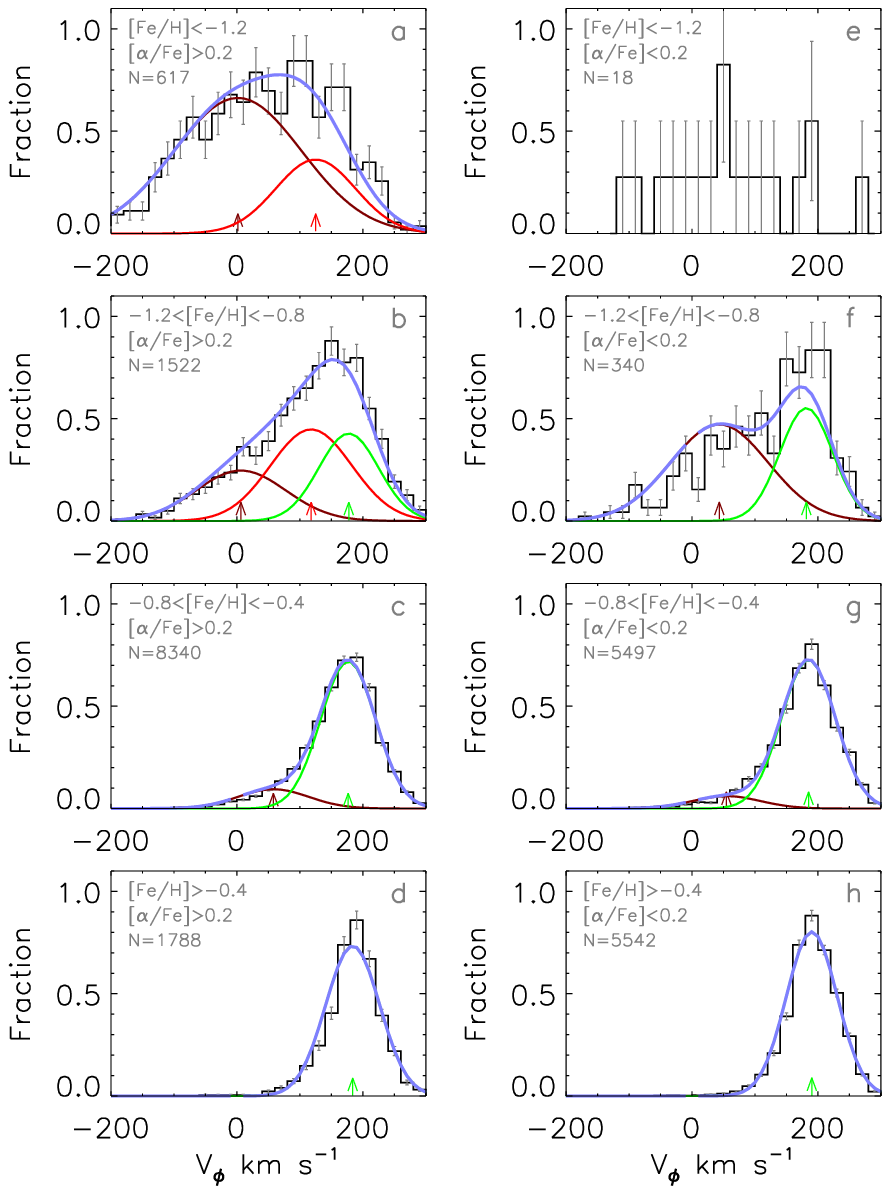}
\caption{Similar to Fig.~\ref{fig:lz_inner} but for the {\VP} distribution.\label{fig:vp_inner}}
\end{figure*}

\begin{figure*}[ht]
\centering
\includegraphics[scale=1.6]{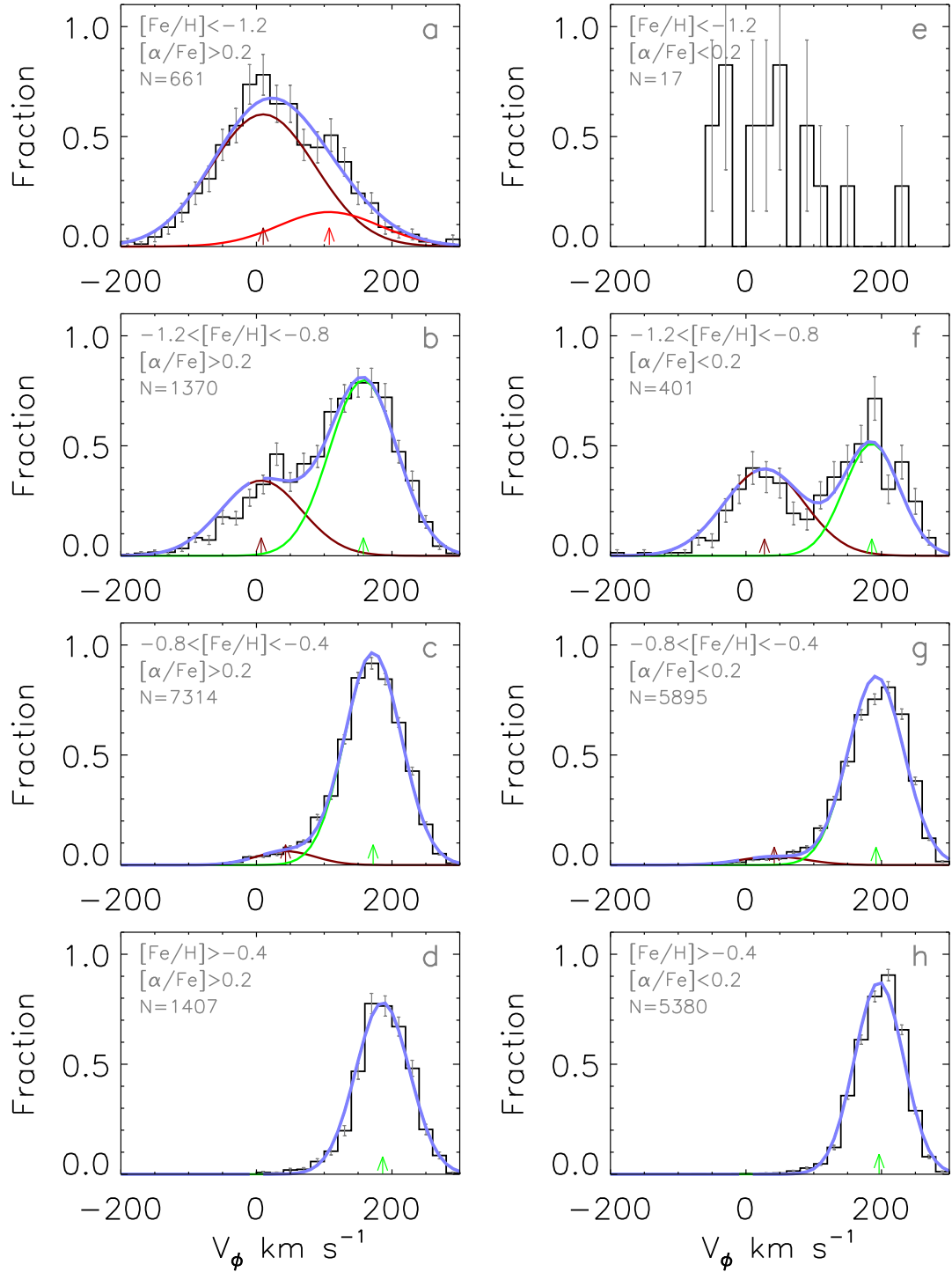}
\caption{Similar to Fig.~\ref{fig:vp_inner} but for the outer stars ({\RGC} $>$ 8.34\,kpc).\label{fig:vp_outer}}
\end{figure*}

It is noted by \citet{An2021}, \citet{Carollo2010} and \citet{Kordopatis2013} that MWTD has kinematic properties different from those of the halo and TD, thus, we display the {\VP} distribution of the inner and outer region stars in Figs.~\ref{fig:vp_inner} and \ref{fig:vp_outer} in slices of [Fe/H] and [$\alpha$/Fe], respectively. The best-fit result obtained with the MCMC simulation are superimposed on the histogram with color curves, and the best-fit parameters are listed in Tables~\ref{tab:vp_inner} and \ref{tab:vp_outer}.

\begin{table*}[]
\tiny
\centering
\setlength{\tabcolsep}{1.5pt}
\caption{The parameters of the best-fit components in the {\VP} distributions for inner region stars ({\RGC} $<$ 8.34\,kpc). The units of $\left \langle {\mathrm{V_{\phi}}} \right \rangle$ and $\sigma_{\mathrm{V\phi}}$ are in km\,s$^{-1}$.\label{tab:vp_inner}}
\begin{tabular}{ccccccccccccccc}
\toprule
\multirow{3}{*}{[$\alpha$/Fe]} & \multirow{3}{*}{Component} & \multicolumn{3}{c}{[Fe/H]$<$$-$1.2} & 
\multicolumn{3}{c}{$-$1.2$<$[Fe/H]$<$$-$0.8} & \multicolumn{3}{c}{$-$0.8$<$[Fe/H]$<$$-$0.4} &
\multicolumn{3}{c}{[Fe/H]$>$$-$0.4} \\
\cmidrule(r){3-5} \cmidrule(r){6-8} \cmidrule(r){9-11} \cmidrule(r){12-14}
 & &  fraction &  $\mathrm{V_{\phi}}$ peak & $\sigma_{\mathrm{V\phi}}$ 
&  fraction & $\mathrm{V_{\phi}}$ peak & $\sigma_{\mathrm{V\phi}}$
&  fraction &  $\mathrm{V_{\phi}}$ peak & $\sigma_{\mathrm{V\phi}}$ 
&  fraction &  $\mathrm{V_{\phi}}$ peak & $\sigma_{\mathrm{V\phi}}$    \\
\midrule
\multirow{3}{*}{$>0.2$} & TD & & &  & 0.38$\pm$0 & 177.4$\pm$7.2 & 47.6$\pm$3.9 & 0.88$\pm$0 & 176.5$\pm$0.9 & 43.6$\pm$0.6 & 1.00$\pm$0 & 183.7$\pm$1.0 & 42.9$\pm$0.7 \\
              & MWTD  & 0.35$\pm$0 & 124.9$\pm$6.3 & 63.4$\pm$6.5 & 0.40$\pm$0.12 & 117.8$\pm$10.7 & 64.6$\pm$8.8 \\
              & the halo  & 0.65$\pm$0.05 & 1.4$\pm$6.8 & 101.6$\pm$5.5 & 0.22$\pm$0.05 & 6.2$\pm$10.6 & 71.2$\pm$6.2 & 0.12$\pm$0.01 & 57.8$\pm$6.0 & 55.9$\pm$2.6 \\
\midrule
\multirow{2}{*}{$<0.2$} & TD & & &  & 0.54$\pm$0 & 181.5$\pm$6.4 & 42.0$\pm$5.3 & 0.92$\pm$0 & 185.0$\pm$0.9 & 43.0$\pm$0.7 & 1.00$\pm$0 & 190.2$\pm$0.5 & 40.4$\pm$0.4 \\
              & the halo & & &  & 0.46$\pm$0.08 & 43.3$\pm$14.9 & 77.4$\pm$8.2 & 0.08$\pm$0.01 & 54.6$\pm$6.4 & 53.7$\pm$3.1 \\
\midrule
\end{tabular}
\end{table*}

\begin{table*}[]
\tiny
\centering
\setlength{\tabcolsep}{1.5pt}
\caption{Similar to Table~\ref{tab:vp_inner} but for the {\VP} distributions of outer region stars ({\RGC} $>$ 8.34\,kpc).\label{tab:vp_outer}}
\begin{tabular}{ccccccccccccccc}
\toprule
\multirow{3}{*}{[$\alpha$/Fe]} & \multirow{3}{*}{Component} & \multicolumn{3}{c}{[Fe/H]$<$$-$1.2} & 
\multicolumn{3}{c}{$-$1.2$<$[Fe/H]$<$$-$0.8} & \multicolumn{3}{c}{$-$0.8$<$[Fe/H]$<$$-$0.4} &
\multicolumn{3}{c}{[Fe/H]$>$$-$0.4} \\
\cmidrule(r){3-5} \cmidrule(r){6-8} \cmidrule(r){9-11} \cmidrule(r){12-14}
 & &  fraction &  $\mathrm{V_{\phi}}$ peak & $\sigma_{\mathrm{V\phi}}$ 
&  fraction & $\mathrm{V_{\phi}}$ peak & $\sigma_{\mathrm{V\phi}}$
&  fraction &  $\mathrm{V_{\phi}}$ peak & $\sigma_{\mathrm{V\phi}}$ 
&  fraction &  $\mathrm{V_{\phi}}$ peak & $\sigma_{\mathrm{V\phi}}$    \\
\midrule
\multirow{3}{*}{$>0.2$} & TD & & &  & 0.70$\pm$0 & 157.7$\pm$1.9 & 49.8$\pm$1.7 & 0.94$\pm$0 & 172.1$\pm$0.7 & 41.6$\pm$0.5 & 1.00$\pm$0 & 186.2$\pm$1.1 & 39.6$\pm$0.8 \\
              & MWTD  & 0.21$\pm$0 & 107.4$\pm$7.1 & 74.3$\pm$20.7 \\
              & the halo  & 0.79$\pm$0.07 & 10.0$\pm$6.9 & 77.2$\pm$4.0 & 0.30$\pm$0.02 & 7.0$\pm$2.8 & 59.5$\pm$3.0 & 0.06$\pm$0.01 & 43.0$\pm$5.5 & 43.0$\pm$2.9 \\
\midrule
\multirow{2}{*}{$<0.2$} & TD & & &  & 0.56$\pm$0 & 185.6$\pm$4.2 & 40.9$\pm$2.9 & 0.96$\pm$0 & 191.9$\pm$0.6 & 40.9$\pm$0.5 & 1.00$\pm$0 & 196.4$\pm$0.5 & 36.1$\pm$0.4 \\
              & the halo & & &  & 0.44$\pm$0.04 & 27.2$\pm$8.5 & 59.2$\pm$6.4 & 0.04$\pm$0 & 41.7$\pm$5.1 & 48.9$\pm$3.2 \\
\midrule
\end{tabular}
\end{table*}

In each panel of the two figures (Figs.~\ref{fig:vp_inner} and~\ref{fig:vp_outer}), the components determined from the {\VP} distribution are very similar to those from the {\LZ} ones. The {\VP} peak 
of MWTD in the panels a and b of the inner region are 124.9$\pm$6.3\,km\,s$^{-1}$ and 117.8$\pm$10.7\,km\,s$^{-1}$, respectively, while it is 107.4$\pm$7.1 km\,s$^{-1}$ in the panel a of the outer region, which are consistent with those of \citet{Carollo2010} and \citet{Kordopatis2013}, however, they are around 15\,km\,s$^{-1}$ lower than those of \citet{An2021}. We note that the matallicites for the sapmle stars  of \citet{An2021} were  estimated from the photometric data, while the {\VP} of their individual stars were computed using photometric distances and proper-motion measurements from Gaia EDR3, and their sample restrict to stars within $\pm30^\circ$ from the Galactic prime meridian. Therefore, the difference in {\VP} may due to the different in sample stars.
\section{Discussion and Conclusions}\label{sect:conclusion}

Based on the combined dataset of LAMOST DR5 and Gaia EDR3, we investigate the kinematic properties of the giant stars in the solar neighbourhood. Our results show that MWTD occupies a significantly fraction with metallicity lower than -0.8 and [$\alpha$/Fe]$>+$0.2 dex. It is found that the fraction of MWTD is significantly higher within the solar orbit radius, while the Z-axis angular momentum is similar for both subsamples divided by the solar orbit radius, 950 and 1000\,kpc\,km\,s$^{-1}$ for inner and outer region, respectively. The high fraction of MWTD in the inner solar orbit radius and higher $\alpha$ element abundances may suggest that the MWTD stars were born in the inner region of the primordial disk, and some of them migrated to larger radii, or MWTD is the result of inside out scenario. However, the number of the MWTD stars is relatively small compared to the TD and halo ones, thus, it is hard 
\textbf{to} make an ascertained estimation of the origin of  MWTD under the current situations. To figure out the origin of MWTD, additional constraints from both \textbf{observations} (such as the abundances of C, N, O, Mg, Si, Ca and Ti, etc) and numerical simulations are necessary.

\section*{ACKNOWLEDGEMENTS}

Our research is supported by National Key R\&D Program of China No.2019YFA0405500, the National Natural Science Foundation of China under grant Nos. 12090040, 12090044, 11833006, 12022304, 
11973052, 11973042, 1214028, U2031143 and U1931102. This work is supported by the Astronomical Big Data Joint Research Center, co-founded by the National Astronomical Observatories, Chinese Academy of Sciences 
and Alibaba Cloud. This work is also partially supported by the Open Project Program of the Key Laboratory of Optical Astronomy, National Astronomical Observatories, Chinese Academy of Sciences. 
Guoshoujing Telescope (the Large Sky Area Multi-Object Fiber Spectroscopic Telescope LAMOST) is a National Major Scientific Project built by the Chinese Academy of Sciences. Funding for the project 
has been provided by the National Development and Reform Commission. LAMOST is operated and managed by the National Astronomical Observatories, Chinese Academy of Sciences. This work has 
made use of data from the European Space Agency (ESA) mission Gaia (https://www. cosmos.esa.int/gaia), processed by the Gaia Data Processing and Analysis Consortium 
(DPAC, https://www. cosmos.esa.int/web/gaia/dpac/consortium). Funding for the DPAC has been provided by national institutions, in particular the institutions participating in the Gaia Multilateral Agreement.

\label{lastpage}

\end{document}